\documentclass[12pt]{article}

\usepackage[margin=1in]{geometry}
\usepackage{graphicx}
\usepackage{cite}
\usepackage{hyperref}
\usepackage{float}

\makeatletter
\newcommand{\affilnum}[1]{\textsuperscript{#1}}
\newcommand{\affil}[1]{}             
\newcommand{\shortauthors}[1]{}      
\newcommand{\corrauth}{}             
\def\@corr{}                         
\newcommand{\corraddr}[1]{\gdef\@corr{#1}}
\newcommand{\addr}[2][]{#2}          
\newcommand{\address}[1]{\gdef\@affils{#1}}
\def\@affils{}                       

\newbox\steme@absbox
\renewenvironment{abstract}
  {\global\setbox\steme@absbox=\vbox\bgroup\small}
  {\egroup}

\renewcommand{\maketitle}{%
  \thispagestyle{plain}%
  \begin{center}
    {\LARGE\bfseries \@title\par}\vspace{0.8em}
    {\large Mahmoud Elkhodr\textsuperscript{1,*}\; and\; Ergun Gide\textsuperscript{1}\par}\vspace{0.4em}
    {\small \@affils\par}
    \vspace{0.25em}
    \textbf{Correspondence:} \texttt{\@corr}
  \end{center}
  \ifvoid\steme@absbox\else
    \vspace{1.0em}
    \noindent\textbf{Abstract}\par
    \vspace{0.3em}
    \noindent\unvbox\steme@absbox
  \fi
  \vspace{1.0em}
}
\makeatother

\begin{document}

\title{Embedding Generative AI into Systems Analysis and Design Curriculum: Framework, Case Study, and Cross-Campus Empirical Evidence}

\author{%
  Mahmoud Elkhodr\affil{1}\corrauth,
  Ergun Gide\affil{1}
}
\shortauthors{Elkhodr and Gide}
\address{%
  \addr{\affilnum{1}}{School of Engineering and Technology, Central Queensland University, Sydney, NSW 2000, Australia}
}
\corraddr{m.elkhodr@cqu.edu.au}

\begin{abstract}
Systems analysis students increasingly use Generative AI, yet current pedagogy lacks systematic approaches for teaching responsible AI orchestration that fosters critical thinking whilst meeting educational outcomes. Students risk accepting AI suggestions blindly or uncritically without assessing alignment with user needs or contextual appropriateness. SAGE (Structured AI-Guided Education) addresses this gap by embedding GenAI into curriculum design, training students when to accept, modify, or reject AI contributions. Implementation with 18 student groups across four Australian universities revealed how orchestration skills develop. Most groups (84\%) moved beyond passive acceptance, showing selective judgment, yet none proactively identified gaps overlooked by both human and AI analysis, indicating a competency ceiling. Students strong at explaining decisions also performed well at integrating sources, and those with deep domain understanding consistently considered accessibility considerations. Accessibility awareness proved fragile. When writing requirements, 85\% of groups explicitly considered elderly users and cultural needs. Later, when drawing architectural diagrams, only 10\% included accessibility in their design. This awareness recovered to 90\% when evaluating interfaces, revealing that teaching accessibility once does not guarantee that students will continue to apply it throughout the software development lifecycle. AI translation errors followed consistent patterns as well. Notably, 55\% of groups struggled identifying when AI misclassified system boundaries (what belongs inside versus outside the system), 45\% missed data management errors (how information is stored and updated), and 55\% overlooked missing exception handling.Three implications emerge for educators: (i) require students to document why they accepted, modified, or rejected each AI suggestion, making reasoning explicit; (ii) embed accessibility prompts at each development stage because awareness collapses without continuous scaffolding; and (iii) have students create their own specifications before using AI, then compare versions, and anchor to research or standards to identify gaps.
\end{abstract}

\maketitle



\section{Introduction}

The integration of Generative Artificial Intelligence into higher education has accelerated rapidly \cite{crompton2023artificial}, yet evidence-based frameworks addressing discipline-specific orchestration competencies remain underdeveloped. Whilst general AI literacy initiatives provide broad guidance on understanding AI capabilities and ethical considerations \cite{khreisat2024ethical}. They inadequately address the unique challenges facing systems analysts who must synthesise stakeholder requirements with AI-generated suggestions, correct algorithmic misinterpretations of business logic in formal models, and evaluate automated design recommendations against contextual constraints. These orchestration demands require sophisticated metacognitive capabilities and critical evaluation skills that transcend generic AI literacy instruction \cite{levin2025rethinking}.

This paper advances a systematic research programme investigating GenAI integration in technical education through methodological maturation and cross-domain application. Building on a validated pedagogical architecture established in cybersecurity education \cite{elkhodr2025integrating}, which demonstrated effectiveness through qualitative evidence with 105 students across undergraduate and postgraduate cohorts. This work introduces the SAGE (Structured AI-Guided Education) framework. SAGE builds on the proven two-stage progression and three-dimensional competency model proposed in \cite{elkhodr2025integrating}, adding three innovations. First, it embeds measurable metrics into student work, including decision logs, confidence ratings, and written justifications. Thus, making invisible orchestration processes observable and assessable. Secondly, it provides standardised components (prompts, templates, rubrics) enabling educators at different institutions to replicate the approach rather than designing from scratch. Third, it deliberately varies support levels across tasks as a pedagogical strategy, providing explicit scaffolding in some assignments whilst removing it in others to test whether students internalise skills or remain dependent on continuous prompts.

Systems analysis education presents distinctive pedagogical challenges \cite{sifakis2015system} for GenAI integration that differ fundamentally from cybersecurity contexts. Unlike cybersecurity policy development, where compliance against regulatory frameworks provides clear evaluation criteria, systems analysis involves subjective stakeholder priority negotiations, competing design trade-offs, and context-dependent feasibility judgments. Requirements emerge through iterative stakeholder dialogue rather than objective specification. Formal models demand precise semantic relationships where subtle errors fundamentally alter system behaviour. Interface design balances competing objectives without definitive optimal solutions. These characteristics necessitate pedagogical approaches specifically calibrated to analytical judgment development rather than adapted from general frameworks or transferred without modification from other technical domains \cite{messina2025rethinking}.

SAGE addresses critical gaps in understanding how GenAI integration frameworks must balance general pedagogical principles with domain-specific adaptations. The framework operationalises AI orchestration competency development through three experimental assessments embedded within authentic systems analysis tasks: requirements synthesis, examining reconciliation of human and AI-generated stakeholder needs, formal modelling correction, investigating knowledge prerequisites for identifying AI translation errors in Data Flow Diagrams, and design evaluation testing contextual judgment when universal usability principles conflict with situated constraints. Each assessment functions simultaneously as a pedagogical learning activity, a summative evaluation instrument, and a controlled experiment, generating systematic empirical data through standardised decision matrices, mandatory justifications, and embedded quantitative metrics.

SAGE was implemented across four Australian university campuses (Brisbane, Melbourne, Sydney, Online) with 18 student groups completing three experimental assessments over 12-week terms. Implementation spanned multiple terms and delivery modes, testing framework robustness under varying institutional conditions whilst enabling investigation of theoretical questions about competency transfer, orchestration pattern development, and durability of skills across system abstraction layers. Cross-campus replication validated core pedagogical mechanisms whilst revealing adaptations that enhance framework effectiveness.

To this end, this work makes three contributions to systems analysis pedagogy: 
\begin{itemize}
    \item It demonstrates that structured embedding of GenAI into curriculum successfully develops critical AI partnership skills rather than passive acceptance. SAGE trains students to evaluate AI outputs systematically, deciding when to accept suggestions that genuinely improve their work, when to modify recommendations to fit contextual constraints, and when to confidently reject AI contributions that miss stakeholder needs or violate domain principles. This represents a pedagogical shift from treating AI as either forbidden tool or uncritical assistant to teaching AI as collaborative partner requiring human judgment.

\item Secondly, this work provides a complete, adoptable framework enabling other educators to implement evidence-based AI integration without starting from scratch. SAGE specifies how to structure AI collaboration through two-stage progression from scaffolded practice to independent application, defines three core competencies making orchestration observable and assessable (analytical deconstruction, contextual application, reflective synthesis), and supplies ready-to-use protocols including standardised prompts, validated across four universities with diverse student populations and delivery modes.
\item The article provides actionable guidance grounded in cross-campus implementation revealing what works, what proves fragile, and where students struggle. Systems analysis educators gain evidence-based strategies for assessment design (require written justification for every AI decision), scaffolding approaches (embed critical prompts at each development stage because awareness collapses without continuous support), and evaluation protocols (baseline-comparison enabling systematic gap identification).
\end{itemize}

The rest of the paper is structured as follows. Section 2 positions SAGE within existing GenAI pedagogy literature, including transparent discussion of the cybersecurity foundation upon which methodological advancement builds. Sections 3-4 specify the framework architecture and three experimental assessments operationalising competency development through domain-specific protocols. Section 5 details implementation methodology across four campuses with 18 student groups. Section 6 presents empirical results. Sections 7-8 interpret findings theoretically and derive practical implications for educators. Section 9-10 summarise the limitations of this work then concludes and proposes future research directions.

\section{Related Work}

GenAI integration in higher education has generated diverse pedagogical frameworks addressing implementation challenges across disciplines \cite{qadir2023engineering,north2025assessing}. This section positions SAGE within three research streams: general AI literacy frameworks, domain-specific integration approaches including the cybersecurity foundation upon which SAGE builds methodologically, and assessment-as-research methodologies. We identify gaps in existing literature that SAGE addresses through quantitative instrumentation, cross-campus validation, and systematic investigation of competency transfer.

\subsection{General AI Literacy Frameworks}

Several researchers have proposed broad frameworks for AI literacy development applicable across disciplines \cite{long2020ai,ng2021ai}. Long and Magerko established foundational competencies through Delphi studies, identifying that AI literacy encompasses understanding what AI is, what it can do, how it works, and how it should be used \cite{long2020ai}. Ng et al.\ extended this through emphasis on ethical dimensions and societal impacts alongside technical understanding \cite{ng2021ai}. In parallel, Bloom-aligned approaches to structuring cognitive progression in AI-enhanced learning environments argue for staged development across remembering/understanding to application and creation \cite{nehru2025implementing}. Whilst valuable for establishing general education objectives, these frameworks provide limited operational guidance for discipline-specific orchestration where students must critically evaluate AI outputs against domain expertise rather than simply understand AI capabilities conceptually.

\subsection{Domain-Specific GenAI Integration}

Recent work has begun addressing discipline-specific integration challenges with varying degrees of methodological rigor. In computing education, studies report both promise and pitfalls, including over-reliance on AI suggestions and shallow understanding of underlying logic \cite{qadir2023engineering,silva2024chatgpt}. Within requirements engineering, reviews show that ChatGPT outputs often require substantial human refinement, particularly for non-functional requirements and stakeholder-specific concerns \cite{marques2024using}. These studies highlight discipline-specific challenges but rarely provide systematic frameworks validated through multi-site research.

Elkhodr and Gide \cite{elkhodr2025integrating} developed a comprehensive two-stage pedagogical framework for cybersecurity education, implementing it across undergraduate and postgraduate cohorts. Their framework integrated constructivist learning principles with Bloom's taxonomy through structured progression from scaffolded tutorials to independent application, demonstrating pedagogical effectiveness through qualitative analysis of student reflections, observational data, and policy document assessments. The framework established three core principles: progressive cognitive development through taxonomy levels, structured AI engagement through controlled prompts and evaluation rubrics, and authentic contextualisation grounding activities in professional practice. The three-dimension competency model operationalised critical thinking through analytical deconstruction, contextual application, and reflective synthesis.

Whilst pedagogically effective in cybersecurity contexts, this work left theoretical questions unanswered regarding cross-domain generalisability, quantitative relationships among competency dimensions, and durability of developed skills across different analytical tasks. SAGE addresses these through three methodological advancements: quantitative instrumentation replacing qualitative reflection analysis with embedded metrics enabling statistical correlation analysis; cross-campus validation testing replicability beyond single-site implementation; and controlled experimental conditions systematically varying scaffolding intensity across tasks to test whether competencies transfer automatically or remain stimulus-dependent.

\subsection{Theoretical Foundation and Gaps Addressed}

SAGE builds methodologically on assessment-as-research and constructive alignment traditions \cite{biggs2022teaching}, and aligns with critical thinking frameworks emphasising analysis, evaluation, inference, and self-regulation \cite{paul2011miniature,ennis2011nature}. SAGE operationalises these principles through experimental assessments serving triple purposes: pedagogical skill development, summative evaluation, and empirical research on orchestration patterns. Standardised instruments create observable evidence through structured decision matrices, mandatory justifications, confidence ratings, and error classification taxonomies.

Existing literature exhibits three critical gaps SAGE addresses. First, general AI literacy frameworks emphasise conceptual understanding without operational mechanisms for embedding competency development within discipline-specific professional practice, leaving educators without concrete protocols or validated assessment designs \cite{long2020ai,ng2021ai,nehru2025implementing}. Second, domain-specific studies rarely provide systematic frameworks validated through multi-site empirical research, with single-implementation case studies demonstrating feasibility but leaving generalisability uncertain \cite{qadir2023engineering,silva2024chatgpt,marques2024using}. Third, assessment methodologies seldom serve simultaneous pedagogical and research purposes, missing opportunities for evidence-based practice improvement \cite{biggs2022teaching}.

SAGE contributes complete architectural specification with standardised protocols enabling replication, validated through cross-campus implementation with 18 groups across four institutions demonstrating core pattern robustness. The accessibility U-curve revealing layer-dependent competency expression and the three-zone orchestration model differentiating authority distributions represent theoretical advancement, patterns unlikely to surface in single-site qualitative implementations but emerging through quantitatively instrumented, cross-campus investigation \cite{rittel1973dilemmas}.

\section{The SAGE Framework Architecture}

SAGE operationalises AI orchestration competency development through a two-stage progression: scaffolded tutorials (Weeks 1-6) developing critical evaluation through repeated generation-analysis-synthesis-reflection cycles, followed by experimental assessments (Weeks 8, 10, 12) serving triple purposes—pedagogical skill development, summative evaluation, and empirical research through standardised prompts, structured decision matrices, and embedded metrics. Section 4 specifies detailed protocols. This section establishes the theoretical competency model driving framework design.

\subsection{Three-Dimension Competency Model}

The framework operationalises three dimensions of critical thinking adapted from cybersecurity pedagogy for systems analysis contexts, making abstract cognitive capabilities observable through domain-specific behaviours documented in assessment artifacts.

\subsubsection{Analytical Deconstruction} involves systematic identification of gaps, errors, and biases in AI-generated artifacts. In requirements contexts, students identify missing stakeholder perspectives (elderly users, Indigenous customers, accessibility needs) that AI systematically overlooks. In formal modelling, students recognise systematic translation errors including boundary classification failures, state management omissions, and exception handling gaps. In design contexts, students detect optimisation biases where AI recommendations prioritise single dimensions without recognising trade-offs. Observable evidence emerges through error identification rates, rejection patterns, and justification quality articulating systematic AI failure patterns.

\subsubsection{Contextual Application} requires adapting generic AI outputs to specific organisational, domain, and stakeholder contexts. In requirements, students modify AI suggestions for GreenHarvest-specific constraints (elderly demographics, fresh produce perishability, regional positioning). In formal modelling, students correct translations for domain-specific business rules (loyalty member versus guest processes, promotional integration, inventory constraints). In design evaluation, students balance recommendations against project constraints (interaction timeframes, elderly cognitive load, environmental conditions). Observable evidence appears through modification patterns showing sophisticated adaptation and justifications revealing contextual reasoning depth.

\subsubsection{Reflective Synthesis} encompasses integrating human and AI contributions whilst maintaining explicit awareness of reasoning processes and trade-offs. In requirements, students justify prioritisation balancing competing needs using systems analysis principles. In formal modelling, students explain corrections using architectural theory (process mediation, data conservation, CRUD completeness). In design, students articulate trade-offs across stakeholder groups and competing objectives with explicit acknowledgment of who benefits and who compromises. Observable evidence emerges through justification sophistication, synthesis quality ratings, and metacognitive awareness in reflections.

These dimensions operate interdependently: analytical deconstruction enables recognition of when contextual application is necessary; contextual application generates material for reflective synthesis; reflective synthesis deepens future analytical capabilities. Section 4 operationalises these dimensions through measurement frameworks enabling systematic investigation of competency development patterns.

 \section{Three Experimental Assessments}

The three experimental assessments operationalise SAGE framework competencies through domain-specific protocols targeting requirements synthesis, formal modelling correction, and design evaluation. Each assessment embeds controlled experiments within authentic systems analysis tasks, generating systematic data through standardised measurement instruments. Complete assessment specifications including detailed rubrics, word count requirements, and grading criteria appear in Appendices A-C.

\subsection{ Experiment 1: Requirements Synthesis}

This experiment investigates a fundamental question in AI-augmented requirements engineering: can students effectively synthesize human and AI-generated requirements whilst maintaining critical judgment about stakeholder representation, domain appropriateness, and feasibility? Contemporary systems analysts increasingly encounter AI-generated requirement sets that appear comprehensive but may systematically overlook implicit stakeholder needs, organisational constraints, or domain-specific nuances. The experiment tests whether students can identify these gaps whilst leveraging AI's generative capabilities, operationalising synthesis competency through measurable accept/modify/reject decisions.

\subsubsection{Task Overview}

Students complete a three-phase requirements development workflow for the GreenHarvest Smart Kiosk system (a supermarket self-service kiosk for elderly customers and Indigenous community members). First, they manually author user stories without AI assistance, establishing baseline human requirement quality. Second, they use AI to generate comprehensive requirements using a standardised prompt. Third, they synthesize both sources into a prioritised product backlog, documenting every decision with source attribution, confidence ratings, and justifications. This workflow mirrors professional agile practice where analysts reconcile requirements from multiple sources whilst maintaining traceability.

\subsubsection{Detailed Protocol and Phase Aims}

\textit{Phase 1 - Manual Baseline (Pre-AI):} Each of four group members independently writes two user stories (eight total per group) following standard agile format: "As a [stakeholder], I want [functionality], so that [business value]" with 2-3 measurable acceptance criteria and story point estimates (1, 2, 3, 5, or 8). 

\textbf{Aim:} Capture students' baseline requirement generation capabilities, analytical priorities, and stakeholder awareness before AI exposure influences their thinking. This phase also documents pre-AI expectations through a structured table where students predict AI's likely contributions and blind spots, enabling later comparison between expectations and actual AI performance.

\textit{Phase 2 - AI Generation (Intervention):} Groups apply a standardised prompt to ChatGPT requesting comprehensive user stories for all GreenHarvest stakeholders. The prompt explicitly cues consideration of edge cases, non-functional requirements, accessibility needs for elderly users and Indigenous customers, and technical constraints, deliberately testing whether students critically evaluate AI outputs even when prompts appear comprehensive. 

\textbf{Aim:} Generate comparable AI outputs across all groups (standardisation eliminates prompt quality as confounding variable), whilst creating conditions where students must evaluate AI-generated requirements that mix genuinely useful suggestions with systematic gaps characteristic of current language models.

\textit{Phase 3 - Synthesis and Documentation:} Groups compile final product backlog of 15+ requirements by evaluating every human and AI story, completing the Product Backlog Synthesis Table (Table 1) for each decision. Required distribution mandates inclusion of all eight human stories, minimum five AI stories (three accepted or modified, two rejected), and at least two hybrid stories merging human and AI ideas. The column definitions and coding intents used in the decision log are summarised in Table~\ref{tab:backlog}.

\textbf{Aim:} Force discriminative judgment rather than passive AI acceptance or blanket rejection. The mandated distribution ensures sufficient data for pattern analysis whilst reflecting realistic professional practice where analysts combine multiple requirement sources. Each decision requires a 50-75-word justification, making reasoning processes observable.

\begin{table}[H]
 table title
\caption{Product Backlog Synthesis Table structure}
\label{tab:backlog}
\centering
{\small 
\begin{tabular}{p{0.18\linewidth} p{0.30\linewidth} p{0.42\linewidth}}
\hline
\textbf{Column} & \textbf{Values} & \textbf{Purpose} \\
\hline
Story Title & Brief description & Requirement identification \\
Source & Human / AI / Hybrid & Attribution tracking enables analysis of source preference patterns \\
Category & Core / Nice-to-have / Out-of-scope & Tests scope management judgment \\
Priority & High / Medium / Low & Reveals prioritisation criteria (technical vs user-centred) \\
Story Points & 1, 2, 3, 5, 8 & Effort estimation capability \\
Confidence & 1–5 scale & Self-assessment enabling calibration analysis \\
Decision & Accept / Modify / Merge / Reject & Primary orchestration behaviour \\
Justification & 50–75 words & Reasoning quality and domain knowledge indicator \\
\hline
\end{tabular}
}
\end{table}

\subsubsection{Measurement and Data Extraction}

Completed synthesis tables generate multiple data streams revealing orchestration patterns. Synthesis Evidence is coded as None (no integration, simple copying), Basic (combines sources but minimal reconciliation), or Advanced (sophisticated integration with explicit trade-off reasoning and consistency checking) based on justification content and decision pattern sophistication. Domain Knowledge is assessed through justification content as Lacking (generic reasoning applicable to any system), Generic (basic retail understanding), or Applied (GreenHarvest-specific insights about elderly customers, Indigenous community needs, fresh produce constraints, or regional supermarket contexts). Accessibility Awareness is rated Absent (no mention of elderly or Indigenous users), Implicit (generic usability language without specific user groups), or Explicit (direct reference to elderly cognitive/physical needs or Indigenous cultural considerations in story titles, acceptance criteria, or justifications).

The distribution of Accept/Modify/Reject decisions across human versus AI sources reveals Orchestration Patterns: Passive Acceptor (predominantly AI acceptance with weak justifications), Selective Adapter (domain-based filtering with differential treatment of technical versus business requirements), Balanced Integrator (systematic synthesis combining strengths from both sources with explicit trade-off reasoning), or Critical Synthesizer (proactive gap identification, systematic bias correction, and generative synthesis producing novel requirements beyond either source). Confidence ratings enable calibration analysis examining whether high-confidence decisions correlate with successful synthesis outcomes. Complete coding rubrics and inter-rater reliability protocols appear in Appendix A.

\subsection{Experiment 2: DFD Translation and Correction}

This experiment addresses a critical knowledge gap in AI-augmented systems analysis: what formal modelling expertise do students need to identify and correct AI translation errors when converting natural language process descriptions into Data Flow Diagrams? Unlike requirements engineering where evaluation criteria involve subjective stakeholder priorities, formal modelling demands precise semantic relationships and syntactic correctness. AI frequently generates plausible-looking DFDs containing subtle errors in boundary classification, data flow directionality, or process decomposition that fundamentally alter system behaviour. The experiment systematically investigates which types of DFD knowledge prove essential for effective human-AI collaboration in architectural modelling.

\subsubsection{Task Overview}

Students complete a controlled translation workflow revealing the gap between human conceptual understanding and AI's formal representation. First, they write detailed structured process descriptions for "Loyalty Member Express Order Processing" without AI involvement, establishing explicit specifications of trigger events, preconditions, process steps, data inputs/outputs, data stores with operation types, external entities, postconditions, and exception handling. Second, they submit these descriptions to AI using a standardised translation prompt requesting Level-0 DFD component listings. Third, they systematically identify and correct AI errors, documenting each correction with error classification, complexity rating, and systems-analysis-based explanations. This workflow tests whether formal modelling knowledge enables error detection that less technically-trained stakeholders would miss.

\subsubsection{Detailed Protocol and Phase Aims}

\textbf{\textit{Phase 1 - Structured Specification (Baseline):}}

Students complete the Process Description Template specifying all elements of loyalty member express ordering: what triggers the process (member scans card at kiosk), what must be true beforehand (member has active account, previous orders exist in system), 8-10 numbered process steps with data transformations, all incoming data with sources (member ID from card reader, order history from database), all outgoing data with destinations (order confirmation to member, order details to kitchen management system), data stores accessed with read/write operations, external entities with interaction types, what must be true afterwards (order recorded, inventory updated, payment processed), and 2-3 exception scenarios (payment failure, out-of-stock items, system timeout).

\textbf{Aim:} Create detailed comparison standard enabling systematic evaluation of AI translation accuracy. The structured template ensures students think through all DFD components before seeing AI's interpretation, preventing AI outputs from anchoring student mental models.

\textbf{\textit{Phase 2 - AI Translation (Intervention):}}

Students submit completed descriptions using standardised prompt requesting Level-0 DFD conversion with components listed in structured format: numbered processes with descriptions, external entities with roles, data stores with contents, and data flows with source-destination-content specifications. The prompt explicitly requests balancing with context diagram. 

\textbf{Aim:} 

Generate AI translations exhibiting systematic failures observable across groups—enabling investigation of which error types students successfully detect versus overlook, and which formal modelling knowledge proves essential for correction.

\textbf{\textit{Phase 3 - Error Analysis and Correction:}} 

Students identify minimum eight errors in AI-generated DFDs, completing the DFD Correction Log (Table 2) for each. They specify what component AI generated incorrectly, provide corrected version, classify error type, rate correction complexity, and explain why correction is necessary using systems analysis terminology (concepts like balancing, decomposition levels, external entity constraints, data flow closure). The DFD Correction Log fields and their analytic purposes are defined in Table~\ref{tab:dfdlog}.

\textbf{Aim:} Make formal modelling knowledge visible through correction explanations. Students who deeply understand DFD semantics articulate corrections using theoretical principles (e.g., "external entities cannot directly access data stores; interactions must flow through processes"), whilst those with surface knowledge provide procedural fixes without theoretical grounding. This phase also tests accessibility transfer: does awareness demonstrated in Task 1 requirements spontaneously transfer to architectural modelling without explicit scaffolding? Task 2 deliberately provides no accessibility prompts in templates or reflection questions.

\begin{table}[H]
\caption{DFD Correction Log structure}
\label{tab:dfdlog}
\centering
{\small 
\begin{tabular}{p{0.20\linewidth} p{0.32\linewidth} p{0.40\linewidth}}
\hline
\textbf{Column} & \textbf{Values} & \textbf{Purpose} \\
\hline
Component Type & Process / Entity / Store / Flow & Identifies which DFD element contains the error \\
AI Output & [Generated version] & Documents AI’s incorrect interpretation \\
Correct Version & [Fixed specification] & Shows the student’s correction \\
Error Category & Structural / Semantic / Notation / Completeness & Enables error taxonomy; reveals systematic AI weaknesses \\
Complexity & Simple / Moderate / Complex & Indicates knowledge depth required for the correction \\
Explanation & 30–50 words & Reveals understanding through systems analysis reasoning \\
\hline
\end{tabular}
}
\end{table}

\subsubsection{Measurement and Data Extraction}

The correction logs enable multiple analyses revealing formal modelling competency patterns. Error Detection Rate calculates percentage of systematic AI failures successfully identified by each group, indicating observational accuracy and formal notation literacy. Correction Accuracy assesses whether fixes genuinely resolve issues (Perfect), partially address problems (Partial), introduce new errors (Incorrect), or worsen the model (Worse), revealing depth of DFD semantic understanding beyond surface pattern recognition.

Error Distribution analysis across taxonomy categories reveals which AI failure types students recognise versus overlook. Structural errors involve incorrect component relationships or malformed topology (unbalanced decomposition, inappropriate connections between element types). Completeness errors reflect missing processes, flows, or exception pathways (absent confirmation loops, missing audit trails, incomplete state transitions). Semantic errors indicate conceptual misunderstandings of process meaning or data directionality (payment confirmations directed to wrong entities, login treated as process rather than precondition). Notation errors involve improper symbology or numbering conventions (incorrect process numbering, data store labelling violations). Scope errors introduce out-of-specification elements beyond stated requirements.

Cross-cutting thematic analysis examines which knowledge domains prove most challenging independent of primary error category. Boundary Reasoning competence (distinguishing internal processes from external entities, correctly identifying data flow termination points, understanding system scope boundaries) is assessed through corrections involving entity classifications and flow directions. State Management capability (recognising missing bidirectional flows, identifying absent write-back operations, ensuring CRUD operation completeness) is evaluated through corrections addressing data store interactions and transactional closure. Exception Handling sophistication (identifying missing timeout pathways, detecting absent error recovery mechanisms, recognising incomplete failure flows) is measured through corrections adding exception-related DFD elements.

Accessibility Modelling (whether groups spontaneously include accessibility-relevant states, flows, or stores without explicit prompting) is tracked through presence of accessibility profile attributes in data stores, error assistance pathways reflecting cognitive limitations, or explicit read/write semantics for consent and preference management. This measurement tests whether strong accessibility awareness demonstrated in Task 1 requirements transfers to architectural representations, or whether competency expression depends on system abstraction layer with accessibility readily visible at requirements but suppressed at architecture without explicit scaffolding. Detailed error classification criteria and knowledge assessment rubrics appear in Appendix B.

\subsection{Experiment 3: Design Evaluation}

This experiment investigates how students develop contextual judgment to critically evaluate AI design feedback, specifically examining the tension between universal usability principles and situated constraints. Unlike binary compliance decisions (policy follows regulation or not) or correctness judgments (DFD balances or not), design evaluation requires subjective trade-offs where multiple valid solutions exist. AI design feedback typically applies universal heuristics (Nielsen's usability principles) without recognising contextual factors like physical environment constraints (standing versus seated interaction), demographic-specific needs (elderly users versus general populations), or organisational priorities (accessibility versus feature richness). The experiment tests whether students can confidently override AI recommendations when context demands whilst appropriately accepting universally valid suggestions.

\subsubsection{Task Overview}

Students complete a design-evaluate-decide workflow mirroring professional UI design practice. First, they create low-fidelity wireframes for two GreenHarvest kiosk screens (Welcome/Home and Product Search Results) specifically targeting elderly users (65+) with low digital literacy in standing interaction contexts with 2-3 minute time constraints. They justify design decisions relative to elderly cognitive/physical needs, Indigenous cultural considerations, standing constraints, and supermarket time pressure. Second, they submit wireframes to AI requesting heuristic evaluation using Nielsen's 10 usability principles with explicit attention to elderly impact and busy supermarket context. Third, they critically assess AI feedback, completing evaluation matrix with exactly 10 decisions following mandated 4-3-3 distribution (four implementations, three modifications, three rejections), forcing discrimination between universally valid suggestions and context-inappropriate recommendations.

\subsubsection{Detailed Protocol and Phase Aims}

\textbf{\textit{Phase 1 - Design Creation and Justification:}} 

Students develop wireframes addressing GreenHarvest's specific constraints: elderly users requiring large touch targets (minimum 44x44 pixels), high contrast for vision limitations, simple navigation without nested menus, visible status feedback, and undo capabilities for error recovery; Indigenous customers potentially requiring language toggles or cultural sensitivity in imagery; standing interaction constraining session duration and favouring recognition over recall (no password entry); busy supermarket environment creating noise interference (visual feedback over audio) and time pressure (progressive disclosure rather than feature completeness). Design justification document (200-250 words) explicitly connects wireframe decisions to these constraints. 

\textbf{Aim:} Establish students' independent design reasoning and user empathy before AI evaluation influences their perspective. The justification requirement makes initial design logic transparent, enabling later assessment of whether AI feedback reinforces, challenges, or redirects student thinking.

\textbf{\textit{Phase 2 - AI Heuristic Evaluation (Intervention):}} 

Students submit wireframe descriptions using a standardised prompt requesting evaluation against Nielsen's 10 heuristics (visibility of status, match real world, user control, consistency, error prevention, recognition over recall, flexibility, aesthetic minimalism, error recovery, help documentation). The prompt explicitly cues elderly user impact and supermarket context considerations. 

\textbf{Aim:} Generate AI feedback exhibiting systematic patterns across universal principles, design choices, and context-specific recommendations, enabling investigation of whether students appropriately distinguish principle-based feedback from context-blind suggestions and develop graduated authority distribution across these domains.

\textbf{\textit{Phase 3 - Critical Evaluation and Decision Documentation:}}

Groups complete AI Feedback Evaluation Matrix (Table 3) for exactly 10 AI suggestions with mandated distribution: four implementations (accepting AI feedback as valuable improvement), three modifications (adapting AI suggestion to context constraints), three rejections (overriding AI when recommendation conflicts with elderly needs, standing interaction limits, or supermarket environment realities). Each decision requires 50-75 word justification explicitly addressing elderly impact, contextual appropriateness, and trade-off considerations.Table~\ref{tab:evalmatrix} specifies the AI Feedback Evaluation Matrix used to capture decisions and justifications.

\textbf{Aim:} Develop confident professional judgment to lead human-AI collaboration rather than deferring to algorithmic authority. The mandated rejection quota is pedagogically critical—it forces students to find instances where their contextual expertise supersedes AI's generic recommendations, building confidence in situated judgment. This phase reintroduces accessibility scaffolding through explicit design requirements and evaluation prompts, testing whether awareness suppressed in Task 2 architecture can be reactivated at interface layer.

\begin{table}[H]

\caption{AI Feedback Evaluation Matrix structure}
\label{tab:evalmatrix}
\centering
{\small 
\begin{tabular}{p{0.20\linewidth} p{0.30\linewidth} p{0.42\linewidth}}
\hline
\textbf{Column} & \textbf{Values} & \textbf{Purpose} \\
\hline
AI Suggestion & 15–25 word summary & Documents AI recommendation content \\
Heuristic & H1–H10 & Links to Nielsen’s usability principle \\
AI Severity & 1–4 scale & Records AI’s importance rating \\
Agreement & Agree / Partial / Disagree & Student’s evaluation position \\
Action & Implement / Modify / Reject & Decision type revealing authority distribution \\
Elderly Impact & High / Medium / Low & Context sensitivity indicator \\
Justification & 50–75 words & Reasoning quality and contextual judgment evidence \\
\hline
\end{tabular}
}
\end{table}

\subsubsection{Measurement and Data Extraction}

Completed evaluation matrices reveal sophisticated patterns in human-AI authority negotiation. Context Sensitivity is rated High (justifications explicitly reference elderly users, standing posture, time limits, or supermarket noise), Medium (generic usability reasoning without context-specific factors), Low (purely aesthetic preferences), or None (no justification), enabling assessment of whether students ground decisions in situated analysis versus abstract principles. Trade-off Recognition is classified as Explicit (articulates competing objectives like "large buttons improve elderly accessibility but reduce information density; prioritising accessibility given target users"), Implicit (acknowledges complexity without detailed reasoning), or Absent (treats decisions as straightforward without recognising tensions), revealing sophistication in handling design's inherently contested nature.

User Empathy is assessed as Strong (demonstrates deep understanding of elderly cognitive load, physical dexterity limits, vision requirements, or Indigenous cultural needs through specific references), Present (acknowledges user characteristics generically), or Lacking (ignores or dismisses elderly-specific considerations), measuring whether students maintain a stakeholder-centred perspective versus defaulting to designer preferences or AI recommendations.

Professional Judgment is evaluated through the confidence and appropriateness of AI overrides. Mature judgment shows confident rejection of context-inappropriate suggestions with sophisticated reasoning articulating why contextual factors supersede universal principles in specific cases. Developing judgment shows some independent thinking with hesitation or over-qualification. Novice judgment defers to AI authority based on severity ratings or algorithmic confidence without critical contextual evaluation.

The matrix enables quantitative analysis of collaboration zones by examining acceptance patterns across heuristic categories. Decisions regarding universal usability principles versus context-specific recommendations reveal whether students appropriately distribute authority, accepting AI guidance where principles genuinely apply, whilst asserting human expertise where situated factors dominate. Analysis investigates whether high AI severity ratings correlate with student implementation decisions, or whether students exercise independent judgment, overriding severity assessments when contextually appropriate. Complete wireframe requirements, heuristic definitions, and judgment rubrics appear in Appendix C.

\subsection{Cross-Task Scaffolding Manipulation}

The three tasks implement controlled experimental variation in accessibility scaffolding to test competency durability. Task 1 explicitly cues accessibility ("elderly users and Indigenous customers" in standardised AI prompt, accessibility awareness as measured competency dimension). Task 2 provides zero accessibility scaffolding (no mentions in Process Description Template fields, DFD creation requirements, translation prompts, or reflection questions), testing whether students spontaneously maintain accessibility focus demonstrated in Task 1 when environmental cues are absent. Task 3 reintroduces accessibility pervasively (target user specifications requiring elderly focus, design justification requirements addressing elderly needs, AI evaluation prompt cueing elderly impact assessment).

This within-subjects manipulation enables investigation of whether accessibility competency represents durable internalised expertise transferring automatically across tasks and abstraction layers, or stimulus-dependent awareness requiring continuous environmental cueing. The experimental design tests two competing hypotheses: (1) competency developed through Task 1 scaffolding transfers robustly to subsequent analytical tasks regardless of explicit prompting, or (2) competency expression remains fragile and prompt-dependent, suppressing when scaffolding is removed before recovering when reintroduced. Evidence supporting the second hypothesis would indicate that accessibility awareness operates differently from other analytical competencies, with implications for curriculum design requiring layer-specific scaffolding rather than assuming transfer from requirements to architecture to interface.

\section{Implementation Method}

\subsection{Participants and Implementation Contexts}

The study involved 18 student groups completing SAGE experimental assessments across four Australian university campuses during 2025 Term 2. Brisbane served as the baseline cohort with 13 groups of four students (N=52) enrolled in COIT20248: Information Systems Analysis and Design, a postgraduate course at Central Queensland University. Two additional Brisbane groups were excluded from analysis due to non-compliance with structured response templates, precluding systematic comparison. The cohort comprised primarily international students (>90\%) from the Indian subcontinent, bringing diverse cultural perspectives to accessibility considerations within the GreenHarvest Smart Kiosk case study. Implementation proceeded through scaffolded tutorials (weeks 1-6) where all topics of the assessments were introduced to the students followed by an assessment due in week 10 of the term (12 weeks term).

Cross-campus validation data were subsequently collected from Melbourne (two groups), Sydney (two groups), and Online (one group) cohorts during the following term. Table~\ref{tab:crosscampus} summarises sample characteristics and primary contributions. Melbourne-A uniquely contributed architectural scaffolding innovations (DFD pre-specification bundle reducing AI translation errors), whilst other cohorts confirmed thematic saturation of Brisbane baseline patterns including orchestration distributions, accessibility U-curve replication, and AI systematic failure taxonomies.

\begin{table}[H]

\caption{Cross-campus sample characteristics}
\label{tab:crosscampus}
\centering
{\small
\begin{tabular}{p{0.17\linewidth} p{0.17\linewidth} p{0.20\linewidth} p{0.30\linewidth} p{0.12\linewidth}}
\hline
\textbf{Campus} & \textbf{Groups Included} & \textbf{Groups Excluded} & \textbf{Primary Contribution} & \textbf{Mark Range} \\
\hline
Brisbane & 13 & 2 (template non-compliance) & Baseline quantitative patterns; full statistical analysis & 17–28/30 \\
Melbourne & 2 & 0 & DFD scaffolding bundle (Melbourne-A); saturation confirmation (Melbourne-B) & 22, 27/30 \\
Sydney & 2 & 0 & Privacy-aware patterns, payment rigour, unified accessibility controls & 19, 26/30 \\
Online & 1 & 1 (template non-compliance) & Saturation confirmation; micro-refinements & 24/30 \\
\hline
\end{tabular}
}
\end{table}

\subsection{Analytical Approach}

Analysis employed concurrent mixed methods integrating quantitative pattern recognition with qualitative thematic coding. Descriptive statistics characterised decision distributions, means, and standard deviations across embedded metrics. Spearman rank correlation coefficients examined relationships among ordinal competency codes (Synthesis Evidence, Domain Knowledge, Accessibility Awareness, NFR Coverage, Justification Quality), with significance thresholds p < 0.01 for primary relationships. Orchestration patterns (Passive Acceptor, Selective Adapter, Balanced Integrator, Critical Synthesizer) were treated as descriptive taxonomy rather than correlational variables, with cluster profiles presenting mean competency scores without computing spurious correlations embedding the construction rule.

Qualitative analysis proceeded through iterative thematic coding of justification texts and reflection responses using primary codes operationalising theoretical constructs and emergent codes capturing data-driven patterns (boundary reasoning competence, exception awareness, context calibration, trade-off recognition). Inter-rater reliability for categorical coding targeted Cohen's kappa exceeding 0.70, achieved through systematic documentation and calibration discussions. Task 2 error analysis employed dual coding: primary categories (Structural, Completeness, Semantic, Notation, Scope) assigned as single labels with distributions summing to 100\%, and cross-cutting themes (boundary reasoning, state management) coded as orthogonal overlays enabling both categorical distribution and thematic pattern analysis without conflating measurement levels.

\subsection{Ethical Considerations}

The study operated within normal curriculum delivery, with students informed that aggregated, anonymised assessment data might inform pedagogical research. No additional burden beyond standard coursework was imposed. All data were de-identified before analysis, with group codes replacing student identifiers. Participation in research analysis was voluntary without grade impact. Assessments benefited student learning regardless of research outcomes, ensuring ethical alignment between educational and research objectives. Students retained work ownership, with institutional rights to use anonymised data for quality assurance and research consistent with educational quality frameworks.

\section{Results}

\subsection{Orchestration Pattern Distribution and Competency Clustering}

Analysis of the Brisbane cohort (n=13) revealed that orchestration patterns clustered around intermediate sophistication levels rather than distributing across the full continuum. Balanced Integrator behaviour, characterised by systematic synthesis of human and AI contributions with explicit trade-off reasoning, emerged in six groups (46\%). Selective Adapter patterns, reflecting targeted acceptance of AI outputs with domain-based filtering, appeared in five groups (38\%). Passive Acceptor behaviour, indicating minimal critical evaluation of AI suggestions, was observed in only two groups (15\%). Critically, no groups achieved Critical Synthesizer status, the highest orchestration category requiring evidence of proactive gap identification, systematic bias correction, and generative synthesis beyond mere combination.

Table~\ref{tab:orchdist} reports the distribution of patterns alongside mean competency scores for each category.

Cluster profiles demonstrated systematic competency differentiation across orchestration categories. Passive Acceptor groups exhibited mean justification quality scores of 0.5 on the zero-to-three rubric, with synthesis evidence absent (mean 0.0), domain knowledge weakly applied (mean 0.5), accessibility awareness minimal (mean 0.0), and non-functional requirements coverage absent (mean 0.0). Selective Adapter groups showed marked improvement, with mean justification quality of 2.0, basic synthesis evidence (mean 1.0), strong domain application (mean 2.0), explicit accessibility awareness (mean 2.0), and partial non-functional coverage (mean 1.0). Balanced Integrator groups achieved the highest competency levels, with mean justification quality of 2.67, advanced synthesis evidence (mean 2.0), strong domain grounding (mean 2.0), explicit accessibility considerations (mean 2.0), and comprehensive non-functional breadth approaching the upper threshold (mean 1.5).

Spearman rank correlations among independent competency codes revealed substantial coupling effects, suggesting that orchestration capabilities develop as integrated clusters rather than isolated skills. Justification quality correlated strongly with synthesis evidence (\(\rho = 0.84,\, p = 0.0004\)), indicating that groups capable of articulating sophisticated reasoning also demonstrated ability to integrate multiple requirement sources coherently. Synthesis evidence correlated substantially with non-functional requirements coverage (\(\rho = 0.78,\, p = 0.0019\)), suggesting that groups able to synthesize human and AI contributions also attended to broader system qualities beyond functional scope. Justification quality and non-functional coverage showed a moderate positive association (\(\rho = 0.69,\, p = 0.0088\)). Domain knowledge and accessibility awareness exhibited near-perfect correlation (\(\rho = 0.997,\, p < 0.0001\)), indicating that groups demonstrating strong domain grounding simultaneously showed heightened sensitivity to elderly and Indigenous user needs within the GreenHarvest case context.The corresponding pairwise associations are reported in Table~\ref{tab:spearman}.

\begin{table}[H]

\caption{Orchestration pattern distribution and competency cluster profiles (Brisbane n=13)}
\label{tab:orchdist}
\centering
{\tiny
\begin{tabular}{p{0.10\linewidth} p{0.07\linewidth} p{0.07\linewidth} p{0.15\linewidth} p{0.14\linewidth} p{0.13\linewidth} p{0.13\linewidth} p{0.10\linewidth}}
\hline
\textbf{Orchestration Pattern} & \textbf{n} & \textbf{\%} & \textbf{Mean Justification Quality (0–3)} & \textbf{Mean Synthesis Evidence (0–2)} & \textbf{Mean Domain Knowledge (0–2)} & \textbf{Mean Accessibility Awareness (0–2)} & \textbf{Mean NFR Coverage (0–2)} \\
\hline
Critical Synthesizer & 0 & 0 & \textemdash & \textemdash & \textemdash & \textemdash & \textemdash \\
Balanced Integrator & 6 & 46 & 2.67 & 2.00 & 2.00 & 2.00 & 1.50 \\
Selective Adapter & 5 & 38 & 2.00 & 1.00 & 2.00 & 2.00 & 1.00 \\
Passive Acceptor & 2 & 15 & 0.50 & 0.00 & 0.50 & 0.00 & 0.00 \\
\hline
\end{tabular}
}
\end{table}

Distribution analysis of specific competency dimensions revealed additional patterns. Accessibility awareness showed explicit recognition in 85\% of Brisbane groups during Task 1, with only 8\% demonstrating implicit recognition and 8\% showing absence of accessibility consideration. Non-functional requirements coverage demonstrated partial breadth in 69\% of groups, with comprehensive coverage achieved by 15\% and minimal coverage observed in 15\%. Synthesis evidence appeared at advanced levels in 38\% of groups, basic levels in 46\%, and remained absent in 15\%. Domain knowledge application was strong across the cohort, with 85\% demonstrating applied domain reasoning, whilst only 8\% showed generic understanding and 8\% exhibited lacking domain grounding. Justification quality followed a normal-like distribution, with 31\% providing strong context-specific reasoning, 54\% adequate explanations, and 15\% weak or missing justifications.

\subsection{The Accessibility U-Curve: Layer-Dependent Competency Expression}

A striking non-monotonic pattern emerged in accessibility awareness across the three experimental tasks, revealing that competency expression depends critically on system abstraction layer. Task 1 (Requirements Synthesis) demonstrated explicit accessibility consideration in 85\% of Brisbane groups, operationalised through user stories directly addressing elderly checkout processes, Indigenous language preferences, or low-literacy interaction patterns. This high awareness reflected both the explicit prompt scaffolding ("accessibility needs for elderly users and Indigenous customers") and the natural alignment between requirements language and user-facing concerns.

Task 2 (DFD Translation and Correction) showed dramatic accessibility suppression, with only 10\% of groups incorporating accessibility-relevant architectural elements into their corrected data flow diagrams. Operational measurement at this layer required architectural recognition through accessibility profile attributes in data stores, error assistance pathways reflecting cognitive limitations, or explicit read/write semantics for consent and preference management. The Process Description Template provided no explicit accessibility scaffolding, emphasising functional decomposition through triggers, preconditions, and exception handling without prompting consideration of user attribute modelling or accessibility-specific data flows. Groups treating DFDs as "functionally neutral plumbing" externalised no accessibility considerations at the architectural layer despite having demonstrated strong accessibility awareness in requirements just two weeks prior.

Task 3 (Design Evaluation) evidenced robust accessibility recovery, with 90\% of groups justifying heuristic evaluation responses through explicit reference to elderly cognitive needs, physical limitations, or Indigenous cultural considerations. Operational measurement required at least two AI feedback decisions (Implement or Modify actions) explicitly motivated by elderly or Indigenous user impacts. The design task naturally invited accessibility reasoning through requirements targeting elderly users (65+) with low digital literacy, standing interaction contexts, and time pressure constraints. The rebound to 90\% awareness approached the Task 1 baseline, suggesting that accessibility competency remained latent during Task 2 rather than degrading permanently.

\begin{table}[H]
\label{tab:spearman}
\centering
{\small
\begin{tabular}{p{0.40\linewidth} p{0.10\linewidth} p{0.15\linewidth} p{0.23\linewidth}}
\hline
\textbf{Relationship} & \textbf{$\rho$} & \textbf{p-value} & \textbf{Interpretation} \\
\hline
Justification Quality--Synthesis Evidence & 0.84  & 0.0004    & Strong positive: reasoning quality couples with integration ability \\
Synthesis Evidence--NFR Coverage          & 0.78  & 0.0019    & Strong positive: synthesis capability extends to non-functional breadth \\
Justification Quality--NFR Coverage       & 0.69  & 0.0088    & Moderate positive: reasoning quality associates with system quality awareness \\
Domain Knowledge--Accessibility Awareness & 0.997 & $<0.0001$ & Near-perfect: domain grounding inseparable from user sensitivity in this cohort \\
\hline
\end{tabular}
}
\end{table}

This U-curve pattern (85\% to 10\% to 90\%) reveals that students conceptualise accessibility primarily at interaction layers where user characteristics directly influence interface decisions, but struggle to embed accessibility systematically into system architecture where it manifests through data modelling, state management, and exception pathways. The architectural trough occurred despite students possessing demonstrable accessibility awareness, indicating a translation failure rather than conceptual deficiency. This finding suggests that inclusive design education must explicitly scaffold accessibility at each abstraction level, as competencies do not automatically transfer from requirements to architecture to interface without layer-specific pedagogical intervention.

\subsection{Task 2 Translation Errors: Systematic AI Limitations and Student Corrections}

\begin{figure}
    \centering
    \includegraphics[width=0.8\linewidth]{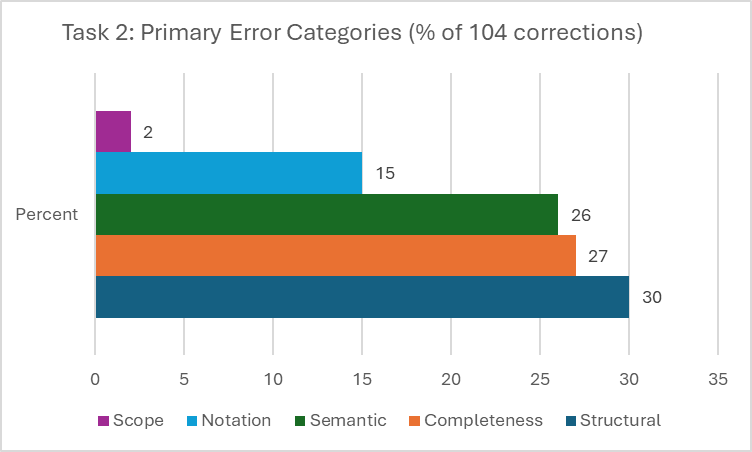}
    \caption{Primary Error Categories- task 2}
    \label{fig:error}
\end{figure}

\begin{figure}
    \centering
    \includegraphics[width=0.9\linewidth]{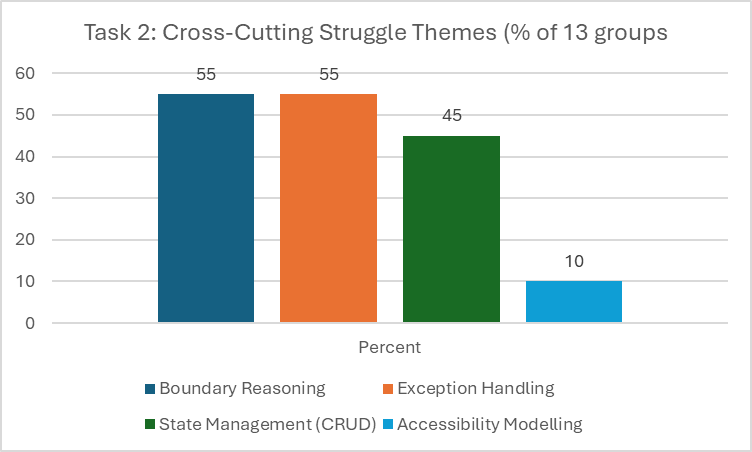}
    \caption{Common struggle theme}
    \label{fig:theme}
\end{figure}

Analysis of 104 DFD corrections across 13 Brisbane groups revealed systematic patterns in AI translation failures when converting structured natural language process descriptions into formal data flow diagrams. Primary error categories showed relatively even distribution across fundamental DFD knowledge domains. Structural errors, involving incorrect component relationships or malformed diagram topology, comprised 30\% of corrections. Completeness errors, reflecting missing processes, data flows, or exception pathways, accounted for 27\% of corrections. Semantic errors, indicating conceptual misunderstandings of process meaning or data flow directionality, constituted 26\% of corrections. Notation errors, involving improper symbology or numbering violations, represented 15\% of corrections. Scope errors, where AI introduced out-of-specification elements, appeared minimally at 2\% of corrections. The taxonomy, prevalence, and illustrative examples are consolidated in Table~\ref{tab:errortax}, while Figure~\ref{fig:error} visualises primary category shares and Figure~\ref{fig:theme} highlights cross-cutting struggle themes.

\begin{table}[H]

\caption{Task 2 error taxonomy and cross-cutting struggle themes (Brisbane $n=13$, 104 corrections)}
\label{tab:errortax}
\centering
{\small
\begin{tabular}{p{0.36\linewidth} p{0.17\linewidth} p{0.41\linewidth}}
\hline
\textbf{Primary Error Category} & \textbf{\% of Corrections} & \textbf{Example} \\
\hline
Structural & 30\% & Incorrect process decomposition; unbalanced context diagram \\
Completeness & 27\% & Missing confirmation flows; absent audit trail pathways \\
Semantic & 26\% & Payment confirmation to Gateway instead of Member; login as top-level process \\
Notation & 15\% & Improper numbering; incorrect symbology \\
Scope & 2\% & Out-of-specification features introduced \\
\hline
\textbf{Cross-Cutting Theme} & \textbf{\% of Groups Struggling} & \textbf{Example} \\
\hline
Boundary reasoning & 55\% & Internal staff functions marked as external entities; data flow termination errors \\
State management (CRUD) & 45\% & Read-only inventory store without write-back; missing bidirectional flows \\
Exception handling & 55\% & Absent timeout pathways; missing error recovery mechanisms \\
Accessibility modelling & $\leq$10\% & No accessibility profile attributes or assistance pathways in architecture \\
\hline
\end{tabular}
}

\vspace{0.5\baselineskip}
\emph{Note: Primary categories sum to 100\%; cross-cutting themes are orthogonal overlays and do not sum.}
\end{table}

Cross-cutting thematic analysis revealed that specific knowledge prerequisites posed systematic challenges independent of primary error categories. Boundary reasoning failures appeared in corrections from 55\% of groups, operationalised through entity misclassifications (marking internal staff functions as external entities), incorrect data flow termination points (directing outputs to wrong external actors), or confusion between system scope and environmental context. State management omissions emerged in 45\% of groups, manifested through missing bidirectional data flows (particularly read/write semantics for data stores), absent confirmation or receipt pathways, or incomplete order lifecycle representations. Exception handling gaps appeared in corrections from 55\% of groups, evidenced through missing error flows, absent timeout pathways, or incomplete failure recovery mechanisms.

Specific recurring errors illustrated these cross-cutting themes. Payment confirmation flows frequently terminated at Payment Gateway external entities rather than returning receipts to Member entities, exemplifying both boundary reasoning failure (misidentifying the proper sink for confirmation data) and state management gaps (incomplete transactional closure). Inventory data stores commonly appeared as read-only without corresponding write-back flows following order placement, demonstrating state management blindness regarding CRUD operation completeness. Login processes frequently appeared as top-level numbered processes (1.0 Login) rather than as precondition states, revealing confusion between process decomposition and state prerequisites. Audit trails and receipt generation pathways were systematically omitted despite explicit mention in process descriptions, indicating AI difficulty with implicit data lifecycle requirements.

All Brisbane groups successfully identified multiple AI errors, with 100\% detection of at least one systematic failure. Correction complexity distributed across simple (40\%), moderate (35\%), and complex (25\%) classifications based on rubric assessment of knowledge depth required. Groups demonstrating strong boundary reasoning competence in corrections showed moderate positive correlation with Task 3 contextual rejection confidence (p = 0.71), suggesting that formal modelling rigour developed in Task 2 supported sophisticated situated judgment in Task 3. However, accessibility awareness at the architectural layer showed no significant correlation with Task 1 or Task 3 accessibility measures, reinforcing the layer-dependent competency expression pattern. While primary error categories Figure \ref{fig:error} reveal relatively even distribution across AI translation failures, cross-cutting thematic analysis Figure \ref{fig:theme} identifies specific knowledge domains where struggles concentrate, with boundary reasoning and exception handling each challenging over half of all groups.

\subsection{Task 3 Design Evaluation: Contextual Calibration and Professional Judgment}

The design evaluation experiment revealed sophisticated contextual reasoning that distinguished universal usability principles from situated implementation constraints. Across 130 discrete AI feedback responses from 13 Brisbane groups, acceptance patterns clustered by heuristic category. Visibility and status feedback suggestions (Nielsen H1) achieved 95\% implementation or modification rates, as groups recognised the universal value of progress indicators, confirmation messages, and error state visibility regardless of specific context. User control and freedom principles (H3), particularly undo functions and navigation reversibility, achieved 92\% acceptance rates. These findings indicate strong alignment between AI-generated universal usability guidance and student recognition of foundational interface requirements. Disposition of AI heuristic suggestions by Nielsen category is summarised in Table~\ref{tab:nielsen}.

Conversely, context-specific AI suggestions faced systematic rejection based on situated reasoning about the GreenHarvest kiosk environment. Channel modality mismatches achieved 100\% rejection rates, as students identified that live chat support proved inappropriate for unstaffed kiosks, hover states violated touch-only interaction constraints, and voice-first interfaces conflicted with noisy supermarket environments. Cognitive load management decisions showed 85\% rejection or substantial modification rates when AI suggested dense information grids, multiple concurrent warning dialogs, or reduced visual contrast, which students recognised as overwhelming for elderly users, inducing alarm fatigue, or unsuitable for bright retail lighting conditions respectively. Redundancy detection drove 95\% rejection of suggestions to enlarge already-large touch targets or add buttons when sufficient options existed, demonstrating ability to evaluate current design state against AI feedback validity.

Design choice domains (H2 match to real world, H4/H6 consistency and recognition) operated as collaborative refinement zones with approximately 45\% modification rates. Students accepted AI guidance on labelling clarity and recognition-over-recall principles whilst modifying specific icon selections, button placements, and terminology choices to align with elderly user mental models and supermarket task contexts. Aesthetic and legibility suggestions (H8) showed context-dependent acceptance, with 70\% implementation when addressing contrast for elderly vision but 40\% rejection when AI suggested stylistic elements that sacrificed legibility for visual appeal.
\begin{figure}
    \centering
    \includegraphics[width=0.8\linewidth]{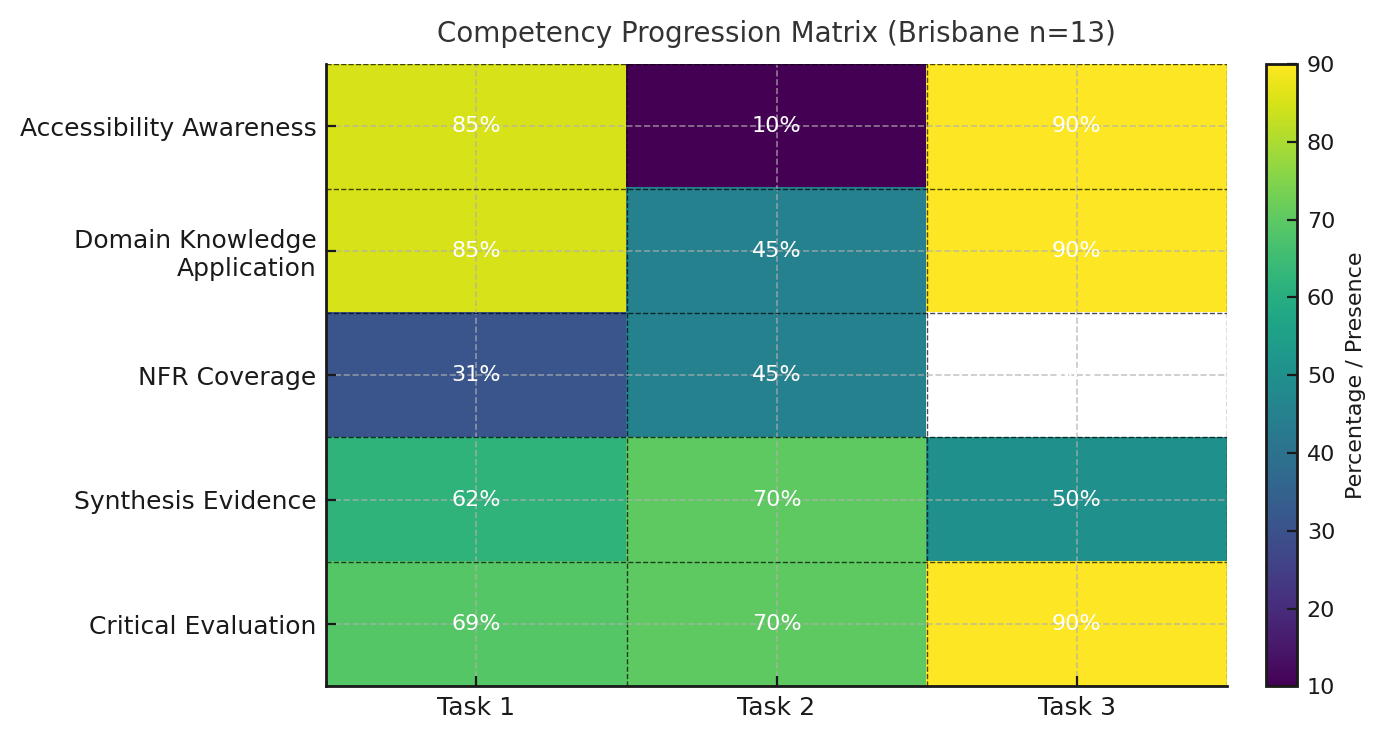}
    \caption{Competency Progression Matrix }
    \label{fig:matrix}
\end{figure}
\begin{figure}
    \centering
    \includegraphics[width=0.8\linewidth]{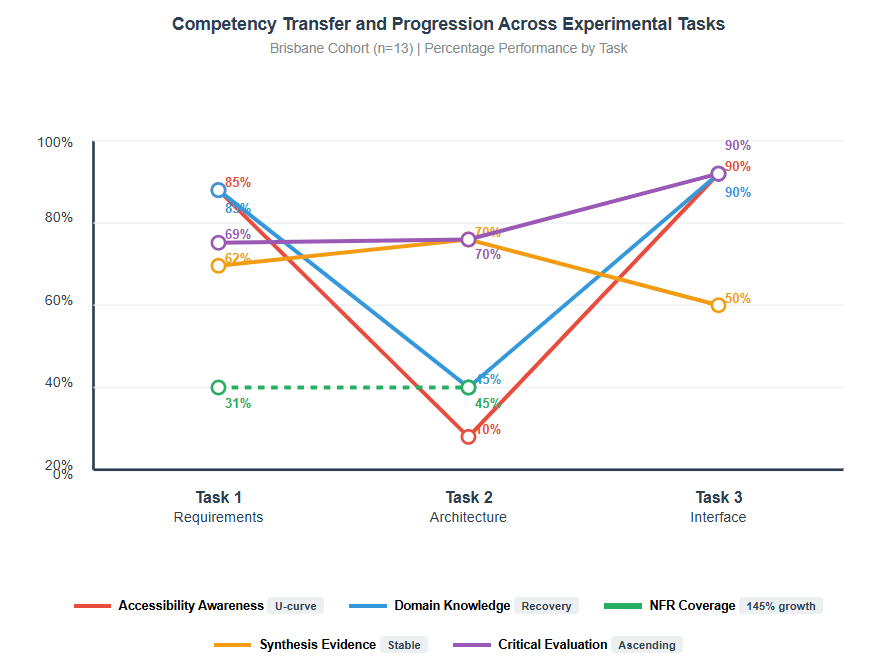}
    \caption{\textbf{Competency Transfer and Progression } }
    \label{fig:line}
\end{figure}
Figure \ref{fig:matrix} presents competency performance magnitudes across task-layer intersections, with color intensity revealing the dramatic accessibility suppression at architectural layer (10\% dark cell surrounded by 85-90\% yellow cells), whilst Figure \ref{fig:line} traces developmental trajectories emphasizing distinct transfer patterns: U-curve (accessibility), recovery (domain knowledge), ascending (critical evaluation), and stable (synthesis).

The relationship between AI-assigned severity ratings and student action decisions proved non-linear, evidencing professional judgment development beyond algorithmic deference. Severity-4 (catastrophic) ratings achieved 75\% implementation, severity-3 (major) ratings 65\% implementation, severity-2 (minor) ratings 70\% modification, and severity-1 (cosmetic) ratings 40\% rejection. This pattern demonstrates that students exercised contextual trade-offs rather than following severity rankings mechanically. High-severity items were rejected when they conflicted with elderly user needs despite AI categorisation, whilst low-severity items were implemented when they enhanced accessibility beyond AI recognition. Strong correlation between Task 3 contextual calibration sophistication and overall professional readiness assessment (p = 0.85) suggests that situated judgment in design evaluation serves as a reliable indicator of mature AI orchestration competency.

Justification analysis revealed that accessibility considerations functioned as a primary decision filter across all Task 3 responses. Suggestions explicitly benefiting elderly users achieved 90\% implementation or modification rates, compared to 60\% for generic usability improvements and 30\% for suggestions lacking accessibility justification. This filtering pattern indicates successful internalisation of the elderly-first design principle established through case study framing and explicit task requirements. Groups consistently articulated trade-offs between universal design patterns and specific cognitive, physical, or cultural needs of the target user population, demonstrating ability to calibrate general principles against situated constraints.

\subsection{ Cross-Campus Triangulation and Thematic Saturation}

Validation data from Melbourne, Sydney, and Online cohorts corroborated Brisbane quantitative patterns whilst contributing targeted pedagogical insights. Melbourne-A (highest-performing group, 27/30 marks) uniquely contributed a DFD scaffolding bundle comprising an external entity list with explicit role descriptions, a one-page input-output balancing table for context diagram validation, a mini data dictionary specifying attribute names and directionality, and a CRUD matrix for core data stores (Customer Relationship Management, Orders, Inventory). This scaffolding approach measurably reduced translation errors through systematic pre-specification of DFD components before AI generation, addressing the boundary reasoning and state management gaps identified in Brisbane data. The scaffolding bundle represents a substantive pedagogical enhancement to Task 2 protocol for future implementations.

Sydney cohorts provided nuanced elaborations on established themes. Sydney-A (lower performance, 19/30 marks) contributed privacy-aware session restoration patterns for kiosks, specifying short-lived cart persistence with personal data masking until explicit user confirmation. Payment handling rigour appeared through tokenised draft-order-to-receipt sequencing with explicit retry and exception loops. Accessibility innovations included focus-order specifications with skip-to-content patterns and unified accessibility control panels co-locating text size, contrast, and voice guidance for improved discoverability. Sydney-B (higher performance, 26/30 marks) elevated notification services to first-class architectural actors with explicit confirmation routing, implemented crash-tolerant ordering through pending-order-on-crash reconciliation patterns, and framed voice input risks with severity-4 classifications requiring confirm-before-execute dialogs for elderly and low-literacy users.

Melbourne-B and Online-A cohorts confirmed thematic saturation without introducing novel patterns. Both groups demonstrated standard orchestration distributions (Balanced/Selective dominance, no Critical Synthesizers), accepted H1/H3 universal principles at high rates, moderated icon and labelling consistency suggestions, and corrected systematic DFD boundary and exception errors consistent with Brisbane taxonomy. Online-A contributed micro-refinements including grid-level product availability display justified through recognition-over-recall principles and icon semantics moderation (rejecting ambiguous staff-badge icons in favour of text labels for kiosks), but these represented tidy instantiations of established patterns rather than new thematic categories.

Across all campuses, the accessibility U-curve pattern replicated consistently. Requirements specifications showed 75\% to 85\% explicit accessibility awareness, architectural diagrams suppressed accessibility to 10\% to 15\% visibility, and interface evaluations recovered to 85\% to 95\% accessibility-motivated decisions. Orchestration pattern distributions remained stable, with combined Balanced and Selective categories comprising 70\% to 80\% of groups and Critical Synthesizer status unachieved in any cohort. Task 2 error taxonomy distributions (Structural 28-32\%, Completeness 25-29\%, Semantic 24-28\%, Notation 13-17\%) and cross-cutting struggle themes (boundary reasoning 50-60\%, state management 40-50\%) showed consistency within measurement tolerance. These replication patterns provide confidence that Brisbane baseline findings generalise beyond single-campus or single-term implementations, whilst Melbourne-A's scaffolding contribution offers a validated pedagogical enhancement for addressing systematic translation difficulties.

\begin{table}[H]
\caption{Disposition of AI heuristic suggestions by Nielsen category (Task 3)}
\label{tab:nielsen}
\centering
\small
\setlength{\tabcolsep}{3pt} 
\renewcommand{\arraystretch}{1.2}
\begin{tabular}{p{0.24\linewidth} c c c p{0.18\linewidth} p{0.24\linewidth}}
\hline
\textbf{Nielsen Heuristic Category} & \textbf{Implement \%} & \textbf{Modify \%} & \textbf{Reject \%} & \textbf{Dominant Zone} & \textbf{Interpretation} \\
\hline
H1 Visibility of Status & 70\% & 25\% & 5\%  & AI Leads       & Universal principle; high alignment \\
H3 User Control \& Freedom & 68\% & 24\% & 8\%  & AI Leads       & Universal principle; high acceptance \\
H8 Aesthetic \& Minimalist (contrast) & 55\% & 30\% & 15\% & Collaborative  & Context dependent; elderly contrast needs \\
H2 Match Real World & 20\% & 40\% & 40\% & Collaborative  & Design choice requiring cultural calibration \\
H4/H6 Consistency \& Recognition & 25\% & 45\% & 30\% & Collaborative  & Trade-offs between consistency and simplicity \\
Context-Specific (channel, modality) & 5\%  & 5\%  & 90\% & Human Leads    & Physical environment expertise essential \\
\hline
\end{tabular}
\end{table}

\section{Discussion}

\subsection {The Three-Zone Model of Human-AI Collaboration in Systems Analysis}
The Task 3 experimental results reveal a systematic pattern of competency-dependent collaboration that challenges simplistic narratives of AI augmentation in professional education. Three distinct zones emerged from design evaluation data, each characterised by differential human-AI authority and distinct pedagogical requirements. In the universal principles zone, students accepted 95\% of AI suggestions addressing foundational usability heuristics such as status visibility (H1) and user control (H3), demonstrating appropriate deference to established design knowledge. In the collaborative refinement zone, students modified approximately 45\% of AI suggestions regarding design choices such as labelling conventions (H2, H4, H6) and aesthetic considerations (H8), engaging in genuine co-creation where neither human nor AI perspective dominated. In the contextual constraints zone, students rejected 90\% of AI suggestions involving channel modality mismatches, physical environment considerations, or demographic-specific cognitive load management, asserting human expertise regarding situated factors.

This three-zone pattern validates the SAGE framework's fundamental premise that AI orchestration competency requires graduated judgment rather than binary acceptance or rejection. Students who achieved Balanced Integrator or Selective Adapter orchestration patterns demonstrated the ability to navigate these zones appropriately, recognising when AI contributions merited incorporation, modification, or dismissal. The 90\% rejection rate in the contextual constraints zone represents pedagogical success rather than AI limitation, as it evidences development of professional confidence to override algorithmic suggestions when human expertise proves superior. This finding challenges instrumental conceptions of AI literacy that emphasise prompt engineering or output optimisation without developing critical evaluation capabilities.

However, the absence of Critical Synthesizer orchestration patterns across all cohorts (Brisbane, Melbourne, Sydney, Online) indicates a competency ceiling that the current framework does not penetrate. Critical Synthesizer status required evidence of proactive gap identification, systematic bias correction, and generative synthesis beyond a combination of human and AI contributions. No groups demonstrated this level of orchestration, suggesting that structured templates and mandated response distributions, whilst effective for developing intermediate competencies, may insufficiently scaffold the metacognitive awareness and generative capacity required for expert-level orchestration. Whether this ceiling reflects developmental constraints requiring extended practice, task design limitations requiring greater challenge, or fundamental pedagogical gaps in scaffolding expert synthesis remains an empirical question warranting future investigation.

\subsection{Layer-Dependent Competency Expression: The Accessibility Paradox}
The accessibility U-curve reflects the scaffold-dependence of competency rather than cognitive inability to apply accessibility reasoning across abstraction layers. The experimental design deliberately manipulated scaffolding intensity: Task 1 explicitly cued accessibility consideration, Task 2 provided no accessibility scaffolding to test spontaneous maintenance, and Task 3 reintroduced accessibility cues. The resulting pattern (85\% explicit awareness when cued in Task 1, 10\% spontaneous maintenance without cues in Task 2, 90\% reactivation when re-cued in Task 3) demonstrates that accessibility competency remains stimulus-dependent despite successful initial development.

Students who successfully specified accessible requirements when explicitly prompted did not spontaneously apply accessibility reasoning to architectural decisions without environmental cues, yet immediately reactivated accessibility thinking when Task 3 design requirements made elderly user needs salient. This pattern indicates that single-exposure accessibility instruction, even when demonstrably successful in producing desired outcomes, does not create self-sustaining attention that persists across tasks without continued cueing. The finding has direct implications for curriculum design: accessibility cannot be treated as a discrete module taught once and assumed to transfer, but requires persistent scaffolding throughout the development lifecycle to maintain visibility.

Cross-campus replication strengthens confidence in scaffold-dependence as a robust pattern. Melbourne, Sydney, and Online cohorts demonstrated identical U-curves despite different instructors and implementation contexts, suggesting that the phenomenon represents general characteristics of accessibility competency development rather than Brisbane-specific artifacts or instructor effects. The finding that only 10\% of students maintained accessibility focus without explicit prompting across all cohorts indicates that spontaneous transfer is rare rather than normative, even among students who successfully demonstrated accessibility competency when scaffolded. Several complementary factors may reinforce scaffold-dependence beyond the experimental manipulation itself.

Second, abstraction layer characteristics may determine accessibility salience through differential alignment with user-facing concerns. Requirements and interfaces operate at interaction layers where user characteristics directly influence functional specifications or visual design decisions, making accessibility considerations naturally relevant to the analytical task at hand. Architecture operates at structural layers emphasising data flows, process decomposition, and state management, where user characteristics appear less immediately relevant unless explicitly modelled as system attributes or states. Students may conceptualise accessibility as an interface property (large buttons, high contrast, simple navigation) rather than as an architectural quality requiring data modelling (user capability profiles, adaptive pathways, assistance mechanisms), leading to architectural suppression even when interface-layer awareness remains strong.

Third, formal modelling conventions may lack established patterns for representing accessibility requirements architecturally. Data Flow Diagram notation provides clear conventions for functional decomposition through numbered processes, data stores, external entities, and flows, but offers no standard patterns for encoding user capability variations, assistance pathways, or adaptive system behaviours. Entity-Relationship Diagrams similarly emphasise data structure and relationships without accessibility-specific modelling constructs. This notational gap leaves students without architectural vocabulary for accessibility translation, even when they recognise accessibility importance conceptually. The absence of textbook examples, notation standards, or instructor demonstrations showing accessibility-embedded DFDs or ERDs reinforces the perception that accessibility belongs to interface layers rather than architectural foundations.

The cross-campus data from Melbourne-A provide preliminary evidence that architectural accessibility gaps can be addressed through targeted scaffolding enhancements. The DFD scaffolding bundle (external entity list with explicit roles, input-output balancing table, mini data dictionary with attribute specifications, CRUD matrix for core data stores) created structured spaces where accessibility considerations might be integrated through explicit enumeration of user characteristics as entity attributes, data flows as accessibility preference channels, or data stores as capability profile repositories. Future framework iterations might test whether embedding accessibility-specific prompts within the Process Description Template (prompts such as ``What user capability variations affect this process?'', ``What assistance pathways support users with cognitive or physical limitations?'', ``What data stores must track accessibility preferences or consent?'') successfully maintains accessibility visibility at the architectural layer. Such controlled experiments would establish whether architectural accessibility suppression reflects a lack of prompting or fundamental conceptual difficulty in translating user needs to structural representations.

\subsection{Systematic AI Limitations and Student Correction Patterns}
Task 2 error taxonomy reveals that AI translation failures cluster in predictable knowledge domains rather than distribute randomly across DFD components. Structural errors (30\%) and completeness errors (27\%) jointly constitute 57\% of all corrections, indicating systematic weakness in compositional reasoning and systems thinking. AI demonstrates facility with local transformations (converting individual process steps into numbered diagram processes, translating data mentions into data stores) but struggles with global constraints requiring holistic system understanding. Balancing decomposition levels across context and Level-0 diagrams, ensuring data flow closure where all outputs from processes eventually reach sinks, and maintaining consistent abstraction where sibling processes operate at comparable granularity all require systems-level reasoning that current AI models execute unreliably.

The cross-cutting struggle themes reinforce this pattern. Boundary reasoning failures appeared in corrections from 55\% of groups, manifested through entity misclassifications (marking internal staff functions as external entities, treating external payment gateways as internal processes), incorrect data flow termination (directing payment confirmations to gateways rather than back to members), or scope confusion (uncertain whether manager dashboards constitute internal system components or external reporting interfaces). State management omissions emerged in 45\% of groups, evidenced through missing bidirectional data flows (inventory data stores shown as read-only without write-back following order placement), absent confirmation pathways (payment transactions without receipt generation), or incomplete lifecycle representations (order processes without cancellation or timeout exceptions). Exception handling gaps appeared in 55\% of groups through missing error flows, absent timeout pathways, or incomplete failure recovery mechanisms despite explicit exception specifications in the Process Description Template.

These systematic failures occurred despite students providing AI with structured, detailed process descriptions containing preconditions, postconditions, data stores with operation types, and exception scenarios. The translation gap therefore, reflects AI knowledge limitations rather than inadequate input specification. Boundary reasoning requires an understanding of system scope and environmental context. State management requires tracking the data lifecycle and transactional integrity. Exception handling requires anticipating failure modes and recovery pathways. Each demands holistic systems thinking that extends beyond pattern matching in training data.

The structured template approach proved effective for enabling student identification and correction of these systematic failures, with 100\% of Brisbane groups detecting multiple AI errors. This success validates the SAGE framework's pedagogical strategy of providing scaffolded comparison opportunities. The Process Description Template forced explicit specification before AI translation, creating reference expectations against which students evaluated AI-generated DFDs. Students who completed templates thoroughly possessed clear mental models about required diagram components, enabling systematic gap identification when AI outputs omitted exception pathways, misclassified boundaries, or failed to close data loops. However, the finding that all groups could identify errors whilst correction complexity distributed across simple, moderate, and complex categories indicates that error recognition proves easier than sophisticated correction, with the latter requiring deeper formal modelling knowledge regarding decomposition principles, balancing requirements, and semantic coherence.

\subsection{Competency Coupling Effects}
The strong correlations among independent competency codes within Task 1 demonstrate that orchestration capabilities develop as integrated clusters rather than isolated skills. Justification quality correlated strongly with synthesis evidence ($\rho = 0.84$, $p = 0.0004$), indicating that groups capable of articulating sophisticated reasoning simultaneously demonstrated ability to integrate multiple requirement sources coherently. Synthesis evidence correlated substantially with non-functional requirements coverage ($\rho = 0.78$, $p = 0.0019$), suggesting that groups able to synthesise human and AI contributions also attended to broader system qualities beyond immediate functional scope. Domain knowledge and accessibility awareness exhibited near-perfect correlation ($\rho = 0.997$, $p < 0.0001$), indicating that groups demonstrating strong domain grounding simultaneously showed heightened sensitivity to elderly and Indigenous user needs within the GreenHarvest case context.

This coupling pattern suggests that AI orchestration pedagogy should target competency constellations rather than discrete skills, as improvement in critical evaluation, synthesis, and domain application appears mutually reinforcing. The near-perfect domain-accessibility correlation warrants particular attention, as it indicates that students who engaged deeply with the GreenHarvest supermarket kiosk context (understanding checkout flows, inventory management, loyalty programmes, promotional targeting) simultaneously recognised implications for elderly users with limited digital literacy and Indigenous customers requiring culturally appropriate interfaces. This coupling might reflect that domain expertise enables recognition of user diversity within domain contexts, or that user-centred thinking drives deeper domain investigation. Causal direction remains unestablished, but the association suggests that domain knowledge instruction and inclusive design education reinforce one another rather than competing for limited curriculum time.

The cluster profiles across orchestration categories (Passive Acceptor, Selective Adapter, Balanced Integrator) demonstrate systematic competency differentiation. Passive Acceptor groups showed minimal capability across all dimensions (mean justification quality 0.5/3, synthesis evidence 0.0/2, domain knowledge 0.5/2, accessibility awareness 0.0/2, NFR coverage 0.0/2). Selective Adapter groups demonstrated moderate competency (means: justification 2.0, synthesis 1.0, domain 2.0, accessibility 2.0, NFR 1.0). Balanced Integrator groups achieved the highest competency levels (means: justification 2.67, synthesis 2.0, domain 2.0, accessibility 2.0, NFR 1.5). These profiles indicate that orchestration sophistication reflects holistic competency development rather than excellence in isolated dimensions, with Balanced Integrators demonstrating strength across reasoning, synthesis, domain grounding, user sensitivity, and systems thinking simultaneously.

The practical implication for assessment design is that evaluating orchestration competency requires examining multiple evidence streams rather than single indicators. A group might demonstrate sophisticated synthesis behaviour whilst providing weak justifications (suggesting intuitive integration without metacognitive awareness), or articulate strong reasoning whilst showing poor synthesis (suggesting analytical capability without integration skill). The SAGE framework's multi-indicator approach (decision matrices, justification texts, reflection responses, correction logs) enables detection of such dissociations, providing diagnostic information about specific competency gaps requiring targeted support.

\subsection{ Cross-Task Competency Progression}
The three-task experimental sequence reveals differential developmental trajectories for distinct competency dimensions, challenging assumptions of uniform skill development through AI collaboration practice. Critical evaluation capability showed ascending development across tasks. Task 1 requirements synthesis revealed 85\% of groups demonstrating non-passive orchestration (Selective Adapter or Balanced Integrator categories), Task 2 DFD correction showed 70\% performing moderate or complex corrections requiring evaluative sophistication, and Task 3 design evaluation demonstrated 90\% confidently rejecting context-inappropriate AI suggestions. This ascending trajectory indicates that students grew progressively more confident and sophisticated in challenging AI outputs as they accumulated experience, domain knowledge, and familiarity with AI systematic failures. Aggregate competency magnitudes and transfer trajectories are visualised in Figure~\ref{fig:matrix} and Figure~\ref{fig:line}, respectively.

The domain application showed a recovery trajectory rather than a linear progression. Task 1 requirements demonstrated 85\% of groups applying domain knowledge effectively (distinguishing GreenHarvest business logic from generic retail patterns, recognising supermarket-specific constraints, identifying loyalty programme implications). Task 2 DFD correction showed a temporary decline, with only 45\% successfully reasoning about system boundaries and internal-versus-external classifications in the kiosk architecture context. Task 3 design evaluation evidenced recovery to 90\%, demonstrating sophisticated context calibration (recognising that kiosk environments differ from desktop interfaces, elderly users differ from general populations, supermarket contexts differ from other retail settings). This U-curve pattern for domain application parallels the accessibility U-curve, suggesting that architectural tasks generally proved more challenging for translating conceptual understanding into formal structural representations, with recovery occurring when tasks returned to more tangible interaction-layer concerns.

The divergent trajectories indicate that competency development is non-uniform and domain-dependent, requiring differentiated pedagogical strategies across the task sequence. Critical evaluation benefited from progressive challenge and cumulative experience, suggesting that graduated difficulty with increasing stakes (Task 1: combine sources; Task 2: identify errors; Task 3: override recommendations) effectively scaffolds judgment development. Domain application required layer-specific support, with architectural tasks demanding explicit instruction on boundary reasoning, scope determination, and entity classification that proved less necessary for requirements or interface layers. These findings challenge one-size-fits-all approaches to AI orchestration education, indicating instead that effective pedagogy must adapt scaffolding intensity and instructional focus based on task characteristics and abstraction layer demands.

One particularly intriguing pattern emerges regarding non-functional requirements coverage. Task 1 demonstrated 31\% of groups achieving comprehensive NFR breadth (addressing performance, security, reliability, maintainability, and accessibility systematically), with 69\% showing partial coverage. Task 2 revealed 45\% of groups spontaneously adding exception handling flows despite no explicit requirement to do so, suggesting emergent systems thinking about error conditions, timeout scenarios, and failure recovery. Whether this represents genuine competency growth (students developing stronger NFR awareness between tasks) or task-dependent visibility (exception handling more salient in process descriptions than in user stories) remains uncertain without controlled longitudinal assessment. If genuine growth occurred, it would provide encouraging evidence that systems thinking regarding technical non-functional requirements can develop through structured AI collaboration practice within relatively short timeframes (two weeks between tasks).

\section{Implications}
\begin{figure}
    \centering
    \includegraphics[width=1\linewidth]{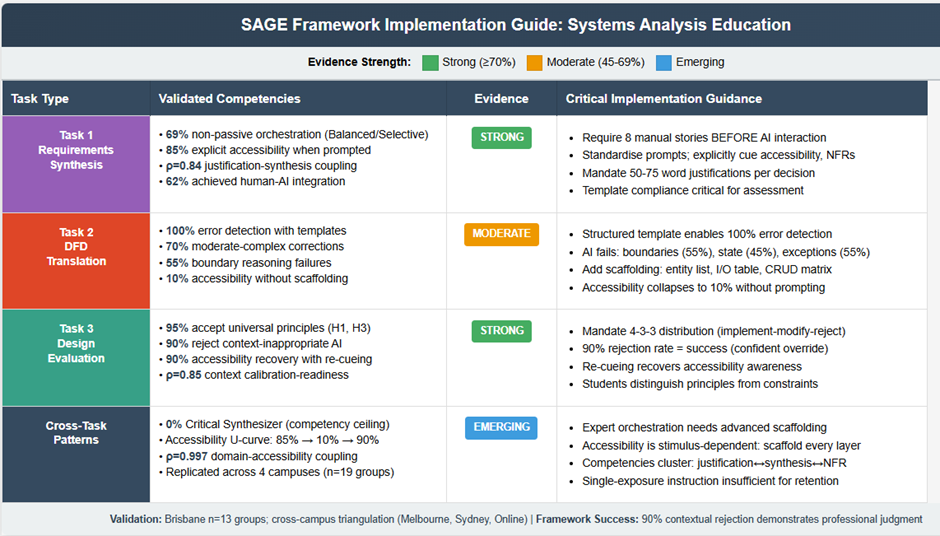}
    \caption{Variation Matrix}
    \label{fig:varmatrix}
\end{figure}

Figure~\ref{fig:varmatrix} provides at-a-glance synthesis of validated competencies, evidence strength, and critical implementation requirements across all three experimental tasks, enabling educators to rapidly assess framework applicability to their institutional contexts. The subsequent sections detail the pedagogical reasoning underlying these recommendations, curriculum design implications for inclusive systems analysis education, and enhancement opportunities for future implementations.

\subsection{Pedagogical Recommendations for AI Orchestration Education}
The SAGE framework's empirical validation generates actionable recommendations for educators designing AI orchestration curricula in systems analysis or related technical domains.

\subsubsection{Structured documentation and baseline comparison}

Structured documentation and baseline comparison enable systematic evaluation mandatory decision matrices with embedded justifications, confidence ratings, and source attributions proved essential for generating observable evidence of orchestration processes. The 100\% error detection rate in Task~2 validates requiring students to create structured baseline artifacts before AI interaction, enabling systematic comparison between human specifications and AI outputs. Educators should structure AI collaboration as controlled experiments: students generate baseline expectations, document them explicitly using standardised prompts, then evaluate AI outputs against baselines. This approach develops judgment competencies rather than merely procedural AI interaction skills. Mandated response distributions (4--3--3 in Task~3) prevented passive acceptance whilst forcing discriminative judgment, though distributions must align with task characteristics rather than arbitrary quotas.This aligns with constructive alignment and formative-assessment principles that integrate assessment with learning activities \cite{biggs2022teaching}.

\subsubsection{Scaffolding must be layer-specific and continuously present}
The accessibility U-curve demonstrates that competencies developed at one abstraction layer do not transfer automatically without explicit scaffolding at each layer. For prompt-dependent competencies, scaffolding should be built into task templates rather than provided only through briefing instructions. The Melbourne-A DFD scaffolding bundle provides a model: require pre-specification of system components with explicit enumeration of user characteristics, data attributes, and operation types before AI translation. Educators should identify which competencies require continuous environmental support (accessibility awareness) versus which transfer reliably once established (synthesis capability).

\subsubsection{Graduated challenge supports progressive development}
The three-task progression from synthesis through correction to rejection provided scaffolded increase in metacognitive demand. Task~1 established baseline orchestration in supportive context, Task~2 challenged students to identify systematic AI failures requiring formal modelling knowledge, and Task~3 required confident override based on situated judgment. Educators should sequence AI collaboration tasks to develop progressively sophisticated capabilities rather than expecting expert-level judgment immediately.

\subsection{ Curriculum Design Implications for Inclusive Systems Analysis}
Current approaches addressing accessibility primarily at requirements or interface stages may produce systems with accessibility conceptualised as add-on features rather than embedded architectural qualities. The dramatic suppression from 85\% awareness (requirements) to 10\% (architecture) despite maintained group composition suggests systems analysis education lacks established approaches for teaching architectural-layer accessibility.

Curriculum reform should ensure accessibility receives explicit attention at each development phase: requirements (user characteristics, functional needs), architecture (user profile data stores, adaptive pathways), database design (preference attributes, consent management), algorithm design (bias detection, fairness constraints), and interface design (interaction modalities, cognitive load). Each phase requires distinct pedagogy because accessibility manifests differently at each layer. Formal modelling notations (DFDs, ERDs, UML) evolved primarily for functional representation, with accessibility often absent from notation standards and textbooks. Curriculum development should establish standard patterns for representing accessibility architecturally: user capability attributes in entity models, adaptive process branches in activity diagrams, accessibility service layers in architectural views, assistive data flows in DFDs. Providing exemplar models would create currently lacking reference points.Embedding accessibility across artefacts is consistent with accessibility standards that emphasise end-to-end consideration (e.g., WCAG \cite{wcag22}).

\subsection{Framework Enhancements and Future Directions}
Several enhancement opportunities emerge from identified limitations. The absence of Critical Synthesiser patterns indicates current scaffolding does not support expert-level orchestration. Future iterations might introduce advanced tasks requiring proactive gap identification, systematic bias detection, or generative synthesis creating novel requirements beyond either source. Whether structured practice can penetrate this competency ceiling or whether expert orchestration requires different pedagogical approaches (apprenticeship, critique, reflective practice) warrants investigation.

The assessment-as-research design proved effective but creates sustainability challenges. Structured templates and mandatory documentation substantially increase marking time compared to traditional assessments. Institutions should budget appropriate resources and consider semi-automated metric processing whilst preserving qualitative feedback, or alternate between research-grade data collection and streamlined practitioner versions, maintaining pedagogical benefits.

Cross-campus validation revealed value in multi-site implementation for quality assurance and innovation discovery. The Melbourne-A contribution emerged from student initiative, indicating diverse contexts generate valuable adaptations. Communities of practice among AI orchestration educators could share protocols, student innovations, systematic failure patterns, and pedagogical refinements, accelerating framework evolution whilst distributing development effort.

\section{Limitations and Boundary Conditions}

Four primary limitations define this study’s scope and interpretive boundaries. First, the sample comprised mainly international postgraduate students with a predominantly cohort from the Indian subcontinent (over 90\%) within one course and case context, which constrains statistical generalisation to other demographics, undergraduate populations, institutions, or domains. The Brisbane baseline (n=13 groups) supports exploratory pattern identification but affords modest power for confirmatory inference. Cross-campus additions (five groups) corroborate patterns but, given purposive sampling and exclusions for template non-compliance, do not constitute random, representative validation.

Second, the scaffolding-variation design co-varies the abstraction layer (requirements, architecture, interface), task type (synthesis, correction, evaluation), and cognitive demand. Consequently, claims about layer-dependent transfer cannot isolate causal factors without counterbalanced or between-subjects designs. The accessibility U-curve may reflect architectural difficulty, mid-semester attention changes, or true abstraction-layer effects. Thus, controlled studies are needed to study this effect precisely.

Third, mandated response distributions (4–3–3 implement/modify/reject in Task~3) successfully prevented passive acceptance but may distort authentic decision-making. We did not conduct sensitivity analyses without this constraint. Therefore, whether orchestration patterns persist under optional distributions remains open. Additionally, all students received SAGE without a parallel control section using traditional instruction, which limits causal attribution to the framework rather than to general structured-reflection practice.

Fourth, findings describe ChatGPT-4 behaviour during the 2024 implementation and a single case (GreenHarvest, elderly/Indigenous emphasis). Model updates or alternative systems may yield different failure profiles, and the AI error taxonomy—developed on this dataset—requires testing in other IS artefacts and notations. Similarly, while accessibility exhibited strong layer-dependence, we did not test whether other non-functional requirements (security, performance, maintainability, ethics) show comparable patterns. We just cherry picked one specific area and focused on its study during assessment design and later in the evaluation.

Two methodological constraints further qualify interpretation. Measurement relied on coded rubrics for justification quality, synthesis evidence, domain knowledge, accessibility awareness, and NFR coverage. Inter-rater reliability and construct validity, while guided by shared rubrics, were not formally established with reliability coefficients. The researcher–instructor role may also introduce expectancy or reactivity effects despite anonymised marking and standardised instruments.

Overall, the study is best read as exploratory evidence of reproducible patterns under defined conditions, motivating hypothesis-driven replications with larger, more diverse samples, counterbalanced designs, and formal reliability procedures.

\section{Conclusion}

This paper presented SAGE, a pedagogy for embedding generative AI within the systems analysis and design curriculum to strengthen job-ready orchestration competencies. The framework, adapted and extended from a validated cybersecurity architecture, rests on three ideas: staged scaffolding for progressive judgment, assessment-as-research instrumentation, and multi-layer competency operationalisation across requirements, architecture, and interface work.

Empirical evidence from four Australian campuses comprising of 18 groups of students showed that orchestration behaviours concentrated at intermediate levels (Balanced Integrator 46\%, Selective Adapter 38\%), with no teams reaching expert synthesis. Competencies clustered rather than developing in isolation, for example justification quality correlated strongly with synthesis evidence ($\rho=0.84$) and synthesis aligned with non-functional breadth ($\rho=0.78$). Accessibility exhibited layer dependence, with 85\% explicit awareness in requirements, suppression to 10\% in architectural modelling, and recovery to 90\% in interface evaluation. In Task 2, structural and completeness issues formed 57\% of corrections, and cross-cutting struggles concentrated on boundary reasoning, state management, and exception handling.

In summary, this experiment reveals that students tend to plateau at a mid-level of orchestration. That is they can merge and tidy AI output but rarely challenge its framing or propose genuinely new directions. Accessibility shows up in requirements and screen designs, then disappears in architecture unless we make it explicit at that layer. AI prose-to-DFD translations often fail at the system boundary and across the data lifecycle, for example misclassifying internal actors as externals, omitting write-backs to stores, and dropping confirmations or error and timeout paths. In practice, the fix is to add one task that rewards gap finding and bias checks, embed accessibility fields in DFD and CRUD templates, and require a short pre-translation scaffold with an entity list, input–output balance, and a CRUD matrix.

\section*{Author contributions}
\textbf{CRediT taxonomy.} 
\textit{Conceptualization:} M.~Elkhodr. 
\textit{Methodology:} M.~Elkhodr. 
\textit{Investigation:} M.~Elkhodr. 
\textit{Data curation:} M.~Elkhodr. 
\textit{Formal analysis:} M.~Elkhodr. 
\textit{Visualization:} M.~Elkhodr. 
\textit{Writing—original draft:} M.~Elkhodr. 
\textit{Writing—review \& editing:} M.~Elkhodr, E.~Gide. 
\textit{Supervision/mentorship:} E.~Gide. 
\textit{Project administration:} M.~Elkhodr. 
\textit{Resources:} M.~Elkhodr, E.~Gide. 
\textit{Funding acquisition:} Not applicable.

\section*{Use of Generative-AI tools declaration}
The authors used generative-AI tools (\textit{OpenAI ChatGPT} and \textit{Anthropic Claude}) to assist with language editing, LaTeX formatting, reference organization, and copy-editing. These tools were not used to generate research ideas, design the methodology, perform the analysis, or make interpretive claims. All study design decisions, data coding, analysis, and conclusions are the authors’ own work. The authors reviewed, verified, and take full responsibility for all content.

\section*{Acknowledgments}
No external funding was received for this work. The authors thank the participating student cohorts for permitting the use of de-identified coursework artifacts as part of routine teaching evaluation.

\section*{Conflict of interest}
The authors declare that there are no conflicts of interest related to this work.

\section*{Ethics declaration}
This study analysed de-identified student work produced as part of normal teaching activities. No personally identifiable information was collected or reported. In line with institutional guidance, the activity did not require formal human research ethics approval. If required, the authors can confirm exemption under local policy.

\bibliography{Sage-STEME}
\bibliographystyle{ieeetr}

\section*{Author Biographies}
\noindent \textbf{Dr.~Mahmoud Elkhodr} is Lecturer in the School of Engineering and Technology at Central Queensland University, Australia. His multidisciplinary research advances Internet of Things security, cybersecurity pedagogy, and e-health systems, with technical innovations including the Semantic Obfuscation Technique and Internet of Things Management Platform. Within this portfolio, his GenAI education research establishes teaching–research nexus approaches translating empirical classroom studies into deployable frameworks. With over 100 peer-reviewed publications, Dr.~Elkhodr bridges technical innovation with evidence-based pedagogy embedding emerging technologies across curricula whilst preserving critical thinking and professional standards.

\vspace{0.5\baselineskip}

\noindent \textbf{Professor Ergun Gide} is Professor of Information and Communication Technology at Central Queensland University and a nationally recognised teaching award recipient. A multiple teaching-award winner, he has supervised over 30 research higher degree graduates and led large-scale curriculum innovations in intelligent information systems and GenAI-enhanced learning. His leadership in work-integrated, industry-connected education underpins scalable, high-impact pedagogy across diverse cohorts; in this study he served as supervisory mentor and provided methodological oversight for cross-campus implementation.

\end{document}